\documentclass[twoside,12pt]{article}
\usepackage{wrapfig,wrapft,graphicx,epsfig}

\def\q{\vec q}

\newcommand {\bea}{\begin{eqnarray}}
\newcommand {\eea}{\end{eqnarray}}
\newcommand {\be}{\begin{eqnarray}}
\newcommand {\ee}{\end{eqnarray}}

\newcommand{\ba}{\begin{array}}
\newcommand{\ea}{\end{array}}

\def\q{{\bf q}}

\def\intd3q{
   \mu^{2\epsilon}\!\!\int {d^{3-2\epsilon}q \over (2 \pi)^{3-2\epsilon}}
}

\def\simge{\mathrel{%
   \rlap{\raise 0.511ex \hbox{$>$}}{\lower 0.511ex \hbox{$\sim$}}}}
\def\simle{\mathrel{
   \rlap{\raise 0.511ex \hbox{$<$}}{\lower 0.511ex \hbox{$\sim$}}}}

\def\N{${\tt N}\,\,$}

\begin{document}
\title{What RHIC Experiments and Theory tell us \\ about Properties of  Quark-Gluon Plasma ? }
\author{Edward Shuryak\\
Department of Physics and  Astronomy,\\
University at Stony Brook, NY 11794, USA}
\maketitle
\begin{abstract}

This brief review summarizes the main experimental discoveries made
at RHIC and then discusses their implications. The
robust collective
flow phenomena are well described by ideal hydrodynamics,
with the Equation of State (EoS) predicted by lattice simulations.
However the transport properties
 turned out to be unexpected, with rescattering cross
section one-to-two orders of magnitude larger than expected from
perturbative
QCD.  These and  other theoretical developments 
indicate that  Quark-Gluon Plasma (QGP)
 produced at RHIC, and probably in a wider
 temperature region $T_c<T<4T_c$, is
 not at all a weakly coupled quasiparticle gas, but is rather
in a strongly coupled regime, sQGP for short.
After reviewing two other ``strongly coupled systems'', 
(i) the strongly coupled supersymmetric
theories studied via Maldacena duality; (ii)
 trapped ultra-cold atoms with very large scattering length,
we return to sQGP and show that there should exist 
literally hundreds of 
bound states in it in the RHIC domain, most them colored.
We then discuss recent ideas of their
effect on the EoS, viscosity and jet quenching. 
 \end{abstract}

  \section{Two sets of major discoveries made at RHIC}

The Relativistic Heavy Ion Collider at Brookhaven is the largest
facility dedicated to heavy ion physics, built 
 {\em to produce and  study the properties of new form of matter},
the Quark-Gluon
Plasma
(QGP). Let me emphasize from the onset that this goal of the
RHIC project has been met with widely spread skepticism,
 especially by people with a high energy physics background.  It was argued
that even if a  large number of quarks and gluons
 be created at RHIC, it will 
 simply fizzle into  a 
firework of multiple jets and mini-jets, with small (and calculable by pQCD)
deviations from a set of
independent multiple
pp collisions. 

RHIC just completed its Run-4, in which a record number of events $\sim
10^9$ per detector was recorded. Obviously our experimental colleagues
are eager to have a look at it now. 
 However, let us all make a break from ever contunuing stream of 
work 
and have a look back, summarizing what have we learned from the data
of Runs 1-3 and recent theory development and comparing it with the
original picture we had in mind many years ago.

 The goal ``to produce QGP'' via certain set of signals,
new flavor, dileptons and phtons, vector meson melting including $J/\psi$,
 together with
the name itself, was first formulated in my papers \cite{Shu_QGP}.
They followed earlier theoretical ideas
  that very high temperature QCD  
should be weakly coupled \cite{CP_75}, in which
the color charge
should not be confined but rather screened \cite{Shu_QGP} (thus ``plasma'').
So QGP was thought to be  a ``normal phase''
of QCD, and expected to be
much simpler in its sturucture than the ``QCD vacuum'', with its
chiral symmetry  breaking and (still mysterious) confinement, leading
to
thousands of quark bound states filling the particle data tables. 
It was widely expected 
 (for about 3 decades!) that a simple perturbative approach
to QGP properties, similar to that used e.g. for QED plasmas,
would  adequately  describe its properties right from $T=T_c$,
at least qualiatively.

But, when one penetrates into
the domain never studied before, one may always find quite
 unexpected things\footnote{Recall that this very country was 
``accidentally'' discovered by Columbus, searching only another
   passage to well known India.
}. The same happened at RHIC. Not only the skeptics
have been proved 
wrong about ``matter'' production and robust collective phenomena
seen there, but the whole view of QGP structure underwent
a major revision in the last year or so.

I will start with a list of  major
discoveries
\footnote{I am certainly not in the position
to  comment on which experiment was the
  first on which particular observations, and use more or less random set
of data, with a reference to the collaboration. (It is possible to use
data from any one of them
due to quite remarkable level of consistency between RHIC data.)
A real experimental summary done by collaborations are expected soon.
}, which I 
 will group 
 into 2 sets,
those related with soft $p_t<2\, GeV$ and hard $p_t>2\, GeV$ 
parts of the observed particle 
spectra.

{\bf Discoveries related with the bulk of secondaries ($p_t<2\,
 GeV$):} are obviously  about the 
  properties of the matter produced. We learned that:\\
 (i) like at CERN,
particle composition is quite well equilibrated, including
 strangeness; \\
(ii) the multiplicity
does not grow very rapidly with energy, as binary scaling for hard
 collisions
 would suggest, so there is some coherencey in production;\\
 (iii)
the magnitude of the radial flow 
velocity is reaching about .7 of speed of light,
 is  larger than at SPS  and its effect extends to higher $p_t\sim
 1.5-2\, GeV$.\\
(iv) Especially impressive are data on the so called
{\em elliptic flow}, observed  for non-central (and even rather peripheral) 
collisions. It is significantly stronger than at CERN.\\
 (v)
Both radial and elliptic flows are correctly described by
hydrodynamics, including their dependence 
on collision energy, centrality and -- last but not least -- the particle type.
It gave good quantitative
 description of about 99\% of the spectra for all
 secondaries\footnote{Modulo the remaining disagreement with HBT
 radii,
on the level of 30-40 percents.}
(except for the hard part at $p_t>2\, GeV$), essentially without any
 parameters
other than (lattice-based) EoS.

  RHIC collisions are 
sometimes called the {\em Little Bangs}, and they are obviously
quite different from a fizzle predicted by pQCD.

{\bf Discoveries  related with the hard tail} of the spectra $p_t>2\, GeV$ , 
 naturally came from runs 2 and 3.\\
 (i)
Already the first data on high transverse momentum tail
of the spectrum, from the second RHIC run,  have shown its  suppression
by a factor $\sim 5$, exceeding expectations of the naive parton
 model. Including such initial state effects as Cronin effect,
one finds that actual jet quenching is closer to a factor 10 suppression.\\
(ii) The puzzle became more intense when
it was found that even the large $p_t$ particles are 
emitted very anisotropically
in the azimuthal angle. \\
(iii) Observation
of the 2-particle correlations at large  $p_t$ have confirmed
that in these region (of not-so-large $p_t$) 
the secondaries still originated from jets.
  The shadowing
of the away-side jets confirmed the strong quenching,
as it also reaches about an order of magnitude suppression
. \\
(iv) Further
 clarification came from the run 3, when  a control experiment
with deuteron-gold (dAu) collisions has confirmed that
at mid-rapidity\footnote{The latest dAu data from BRAMS experiment
indicate much stronger initial state effects in the forward
region, hopefully an expected gluon saturation signal.
} the suppression is $not$ due to the initial state shadowing,
but is instead indeed a final state absorption.

The original intention \cite{quenching} 
was to use   jet  
 quenching  in order to get a kind of a  tomographic picture
 of the QGP cluster. However,  the data available so far (up to 
 about $p_t=12 \,GeV$  mostly show that a produced matter 
 is so black -- up to 90\% of produced jets are absorbed -- that
only jets from  
the surface is escaping. Clearly a drive to larger $p_t$ is still
very much needed.

In this brief report I would not discuss the large $p_t$ physics,
except of mentioning some new ideas about the mechanism of jet quenching, but 
will try to summarize in more detail the physics implications
of the data at smaller $p_t$ on the QGP production and properties.
Before we go into specifics, let me summarize the successes and surprises
we have seen on the way.

Brief summary of QGP properties, as it is extracted from data: 
The thermodynamics at chemical freezeout tells us that
 it occurs
at a universal  temperature
$T_c\approx 170 \, MeV$ which coincides with the expected critical
temperature.
The hydrodynamics  tells us parameters of the EoS: it has not yet been
very precisely mapped (we need an energy scan for that), but 
e.g. the {\em latent heat} of the QCD transition is
fixed by these works to be about 800 $MeV/fm^3$,
 the same value as it was predicted by the lattice QCD.
Furthermore, the  expected EoS
(pressure as a function of the energy density)  above the transition
region is also confirmed to be roughly 
$p\approx \epsilon/3$.

 In contrast to that,
 the transport properties (viscosity) of QGP turned out to be
completely unexpected.  The rescattering of constituents
needed to sustain the observed degree of collectivity
is one to two orders stronger than it was predicted on the basis of
pQCD. 
The ratio of the QGP viscosity relative to its entropy density
$\eta/s\sim 1/10$, making it {\em the most ideal fluid
ever observed}. (In particularly: water would not flow, if only
few thousends molecules would be put together.)

\section{Evolving theoretical views on QGP properties}

Since 1970's till quite recently 
 Quark Gluon Plasma (QGP) was viewed as a gas
 of quasiparticles (dressed quarks and gluons) which  interact relatively
weakly with each other. 
A significant amount of 
theoretical work has been invested on refining the
perturbative high-T
calculations, to thermodynamics and kinetics
of QGP, see e.g. my book \cite{Shu_book2} for more details.
 We now know all  perturbatively calculable
corrections to free gas expressions, 
$O(g^2,g^3,g^4,g^5,g^6log(g))$, making in total 7 terms
of the weak coupling series (see e.g. \cite{Kajantie}). 
 Although they are not
 converging, unless $T\sim 10^6 GeV$ or so,  hopes 
remained that  some clever
re-summation will get all the physics right.

\begin{figure}
\centering
\includegraphics[width=8.cm]{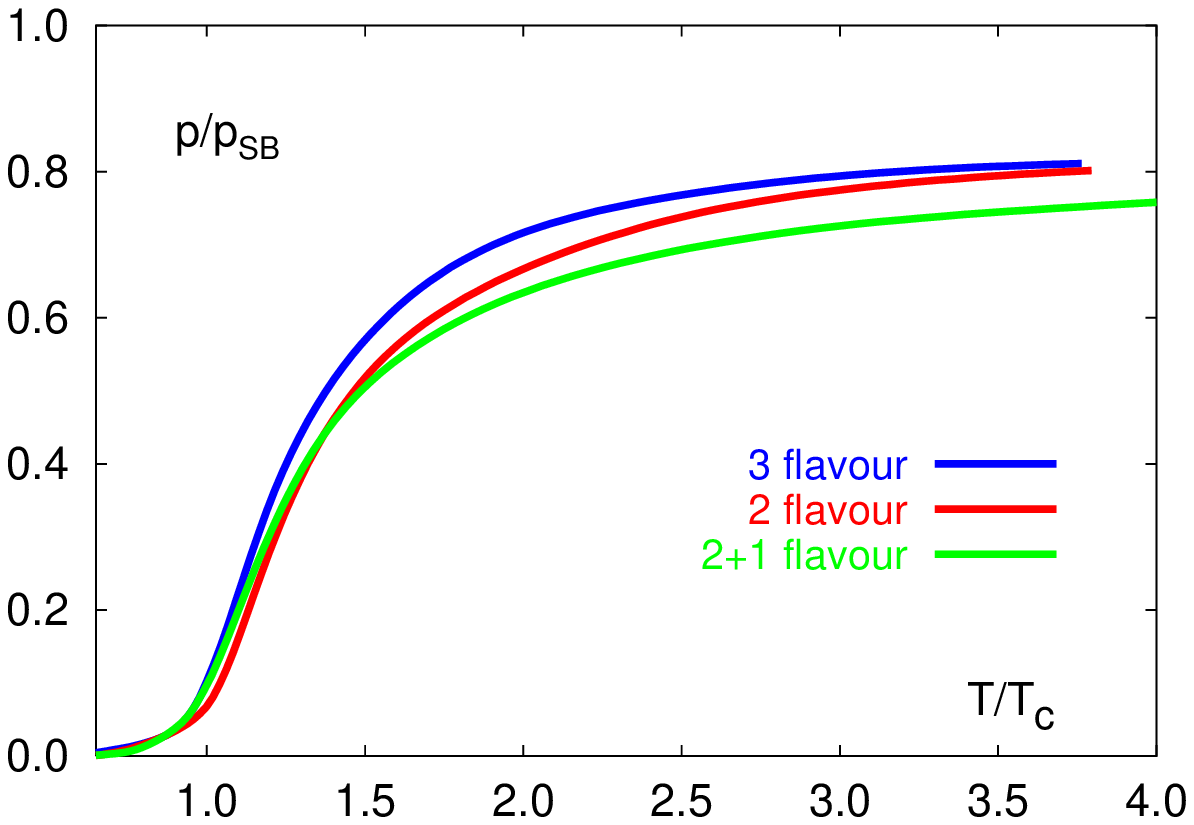} 
      \includegraphics[width=8.cm]{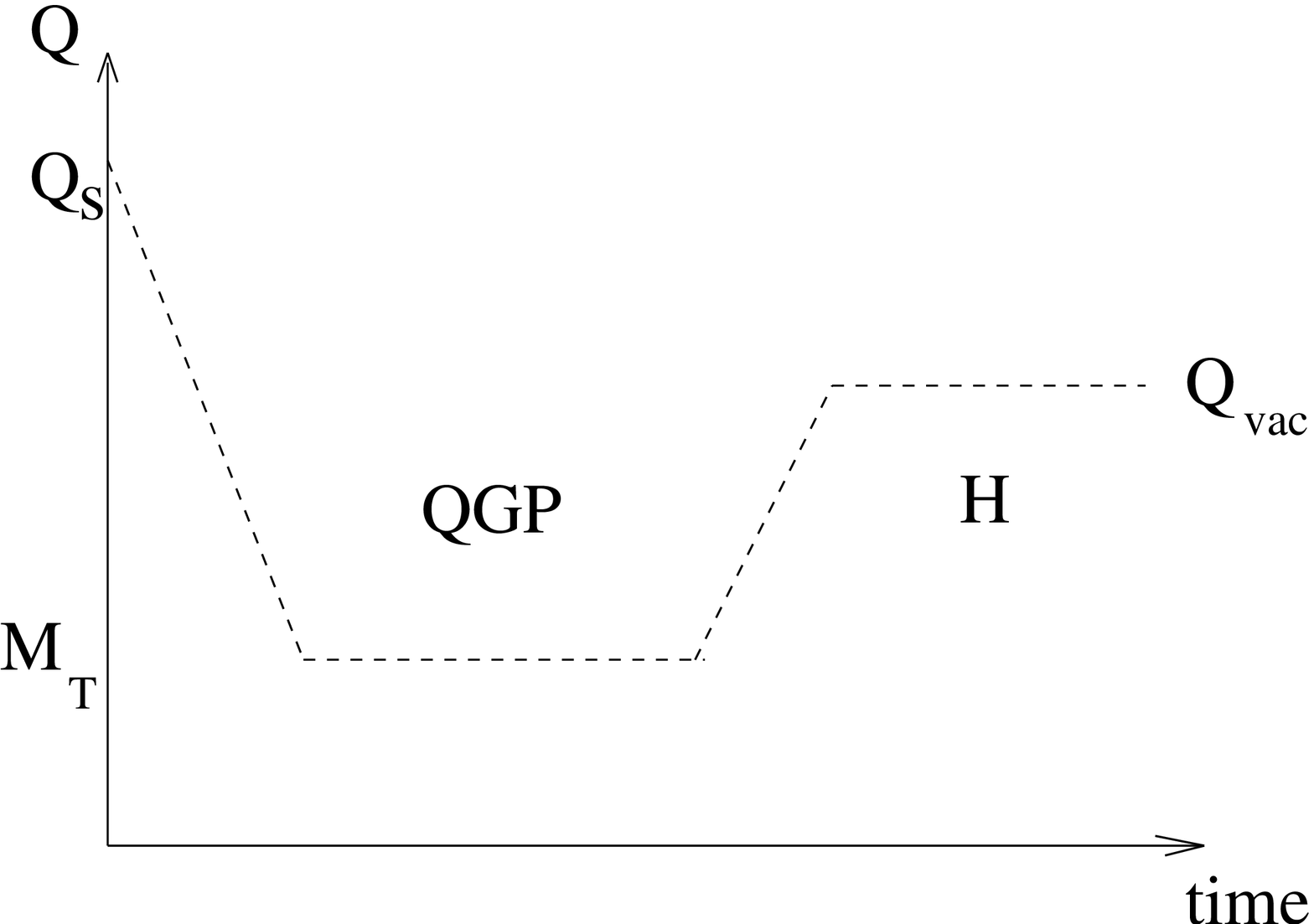}
  \caption{\label{fig_pressure}
(a) The pressure (divided by that for free gas)  versus
the temperature $T/T_c$, from lattice  
thermodynamics studied by Bielefeld group.
(b)
Schematic plot of the cut-off scales during the evolution
of the system with time, from \cite{Shu_spha}. 
At the collision time=0 the scale
is presumably the saturation scale $Q_s$ in the incoming nuclei,
which grows with the collision energy. Then the cutoff decreases
reaching some nearly constant value in QGP at $T>T_c$,
the thermal gluon mass $M_T$ (\ref{eqn_thermal})
and stay at this value till it rises again in the mixed phase to its
vacuum value in the hadronic (H) phase $Q_{vac}\sim 1 \, GeV$.
 }
\end{figure}


The
non-perturbative results which came from lattice QCD, such as pressure
shown in Fig.\ref{fig_pressure}(a), seemed to support this view. 
Indeed, the ratio of the calculated pressure to that at zero coupling
gets to about 0.8 soon after the phase transition, so it was tempting
to assume that the deviation, $\sim .2$,
is a good measure of the interaction corrections.

As I will argue below in more detail, this simple reasoning
turned out to be very misleading, and it managed to fool us all for decades.
Only recently had we learned that in two other examples of very
strongly interaction matter -- (i) the CFT gauge theory at comparable coupling; and (ii) trapped atoms
at the Feshbach resonances -- the same ratio also about .8.
 (And I will argue by the end of this paper, one 
can explain this .8 for QGP at $T=2T_c$ in a radically new picture,
 with only half of pressure
 coming from a quasiparticle gas and the rest
 from {\em hundreds} of bound states.)

The breakthrough in our thinking \cite{SZ_newqgp}
came basically from 3 different sources:\\
 (i) a general idea
that while confinement and chiral breaking disappear at $T>T_c$ 
and there is not much free charge to generate large Debye mass yet,
the sizes of states and scale at which the coupling is defined 
showld be rather low at $T=(1-2) T_c$, see e.g. Fig.\ref{fig_pressure}
(b).\\
(ii) Triumph of hydro\footnote{
In fact 
the  hydrodynamics with its
 ``non-ideal'' expansion in powers of 
the mean free 
path  is 
 the oldest example of a {\em strong coupling} expansion,
 including the $inverse$ powers of the  cross section.}
, compared with apparent failure 
of weak coupling
``parton cascades'' of various kinds.
 For example, Gyulassy and Molnar
\cite{GM} concluded that the elliptic flow can only by reproduced
by a gluon cascade if the cross section be enhanced by a factor 
of about 50. That forced us to think hard
{\em Why is the Quark-Gluon Plasma at RHIC such
  an ideal fluid ?} Important paper
by Policastro, Son and Starinets \cite{PSS} was a radical step in this direction. 
\\
(iii) The last
 ringing bell came from the lattice practitioners. Surprisingly
to all,  recent works  in Japan  and Bielefeld \cite{charmonium}  
have  found that the lowest
charmonium states are not melting at $T_c$, as was believed previously, but
actually persist to at least $T=2T_c$. Then came
similar evidences that mesonic bound states made of light quarks 
survive we into the QGP phase as
well
\cite{Karsch:2002wv}. 

All these developments provided a hint, that the QGP quasiparticles
 at $T\sim few\, T_c$ 
have much stronger interaction than previously expected,
we have found a {\em strongly coupled QGP}.
 
After more details about hydro and brief discussion of two other 
strongly coupled examples
we will return below to  
recent attempts to understand what exactly sQGP is.

\section{ Collective flows, EoS and transport properties  }

 Is hadronic matter really produced in heavy ion collisions?
Is there something qualitatively
  new in AA collisions, never
 seen in  ``elementary''\footnote{Apart of the large-$p_t$
tail, described by the parton model plus pQCD corrections, it is very far
from being elementary and is very poorly understood. One may argue that
heavy ion collisions, described well by hydro/thermodynamics, are in fact
even much
simpler. }
 pp or $e+e-$ collisions?

  Indeed, the original motivation for
 heavy ion program is not just increase the number of
 secondary particles 
produced per event (up to several thousands at RHIC), but to reach
 a {\em qualitatively different} dynamical regime.   
characterized by a small
 microscopic scale $l$ (e.g. mean free path) as compared to
the macro scale $L$ (the system's size):
$ l \ll L $.
 If this is achieved,
 the fireball produced in
heavy ion collisions should be treated as a macroscopic
body, with thermo and hydrodynamics.

  Statistical models do indeed work remarkably well for heavy ion
 collisions, even at energies lower than RHIC. 
 But they also
work for  $pp$ or $e^+e^-$ (and we still do not know why). 
end to it. 
In contrast to that,  $pp$ or $e^+e^-$ show
no sign of flow effects, see early attempts to see them
 \cite{SZ_radialflow}. So,
 a multi-body
excited systems produced in these case are $not$ macroscopically large.
It is not ``matter'' but just a bunch of  particles.

\subsection{Transverse flow }
Heavy ion collisions, on the other hand, showed 
 a variety of 
  {\em``flows"} since very low energies, but not
all of them are indeed 
 a collective expansion. 

    Let me start with a historic comment.
 First attempts to connect the experimental information with
   the collective transverse flow  were
 made independently  by Siemens and
   Rasmussen
\cite{SieRas} for low energy (BEVALAC) and by 
Zhirov  and  myself \cite{SZ_radialflow} for high energy pp collisions
at CERN ISR around 1979.
The idea of both papers was exactly the same: 
 collective expansion boosts 
    spectra  of various secondaries  differently, depending on their mass. 
 Pions are light
and their  spectrum remains exponential,    with  a
 ``blue shifted''  temperature, while for
    heavy particles the effect  is  
 different. Fortunately it is  easily calculable and depends on 
the value of the particle mass only.

 Siemens and Rasmussen found  
  the expected difference for  pions  and
  protons produced  by heavy ions at  $E\sim 1 GeV/N$  fitted  them with two
  parameters, the freezeout temperature $T_f\sim 30 \, MeV$
and the velocity of what they have called the  "blast  wave".
(Long discussion afterwords shown that ideal hydro is not really
  applicable in this case,
however.). Zhirov and myself
 \cite{SZ_radialflow}  found no flow in $pp$ data from ISR:
the  $\pi,K.N$ spectra from $pp$ showed the same $m_t$-slope.
 All of us had to wait for heavy ion
collisions at high energies, and only at SPS and
now at RHIC we have seen
 real hydrodynamical flow, radial and especially elliptic.

\begin{figure}[t]
  \includegraphics[width=10.cm]{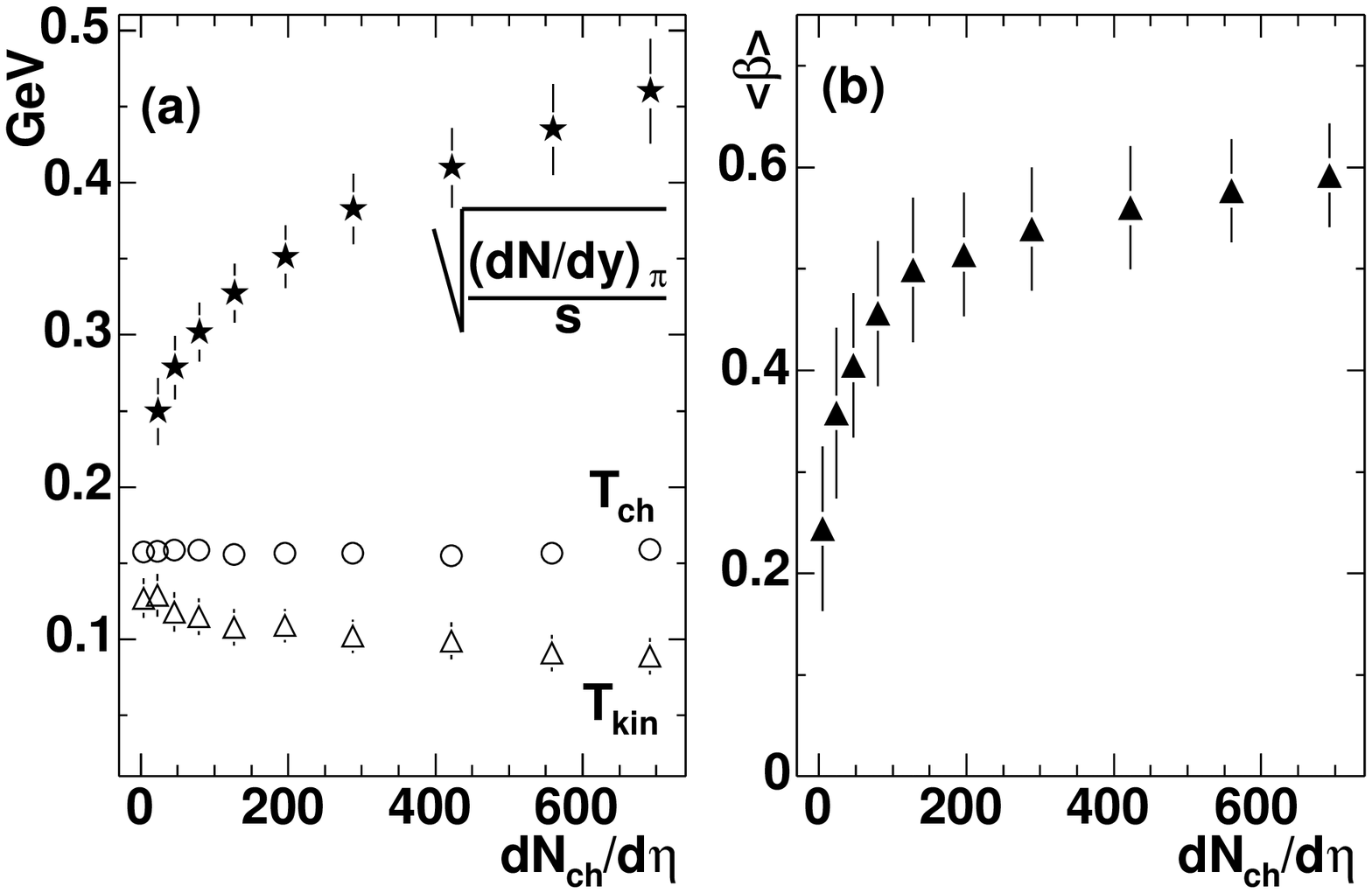} 
 \includegraphics[height=6.0cm, width=6.0cm]{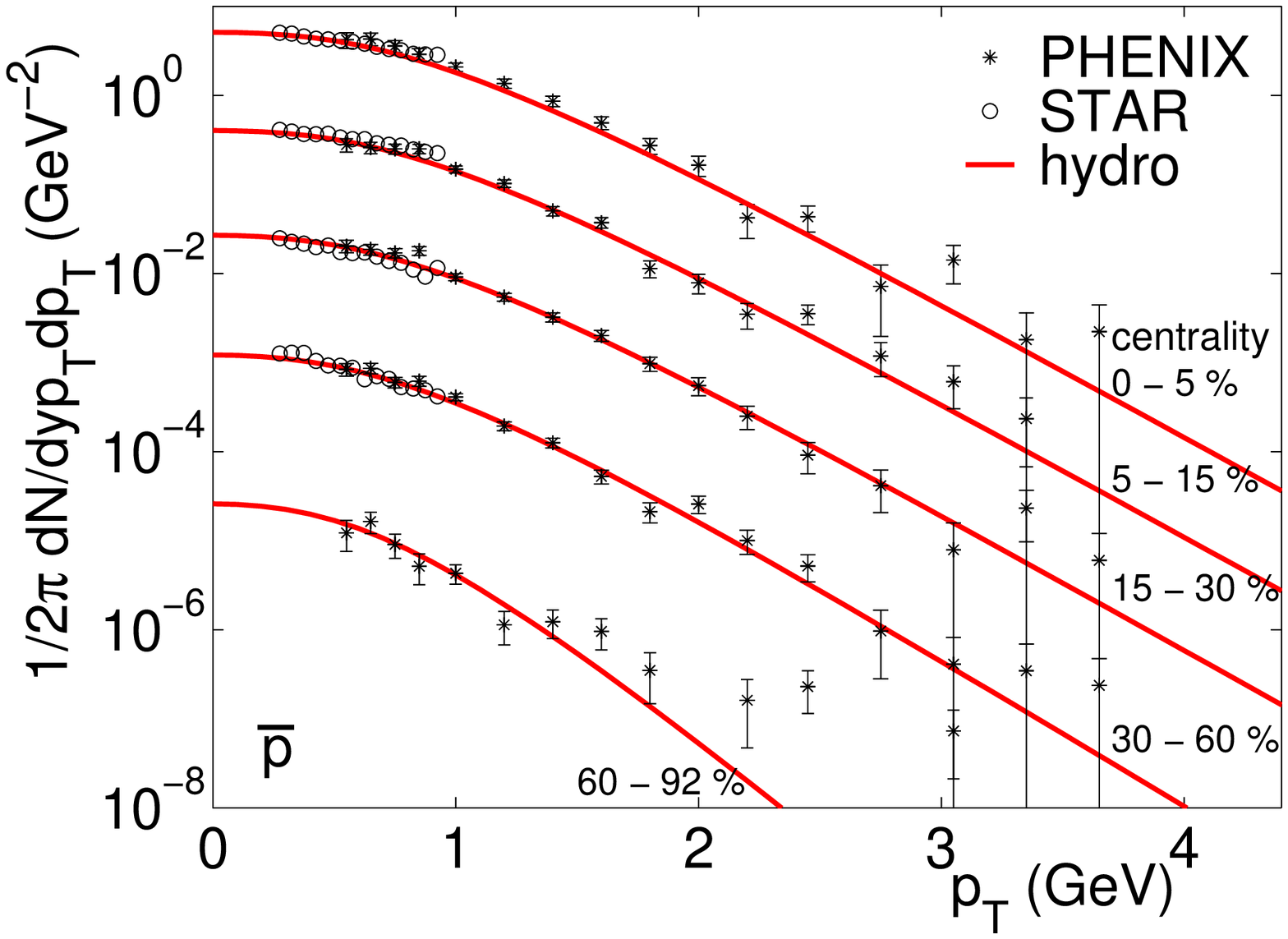}
 \caption{  \label{fig_mtspec_all}
(a) The ``blast model'' fits to STAR collaboration data.
 The values of the  
and freezeout temperatures are shown in (a) and the mean collective
 velocity.
(b) Comparison between STAR and PHENIX data for protons with hydro
calculation by Kolb and Rapp \cite{Kolb:2002ve} (which correctly incorporates
chemical freezeout).
}
\end{figure}

   Let me  jump years ahead and show a modern version of
the blust wave fit to RHIC data, shown 
in Fig.\ref{fig_mtspec_all} as a function of centrality.
Two basic parameters are the freezeout
temperature $T_{kin}$ and the mean flow velocity $<\beta>$. 
 $T_{kin}$ decreases and the velocity increases for more central collisions, 
displaying an expected
 conversion of the internal energy into flow. 
Note also that the temperature of chemical equilibration $T_{ch}$ 
seem to be completely independent of the 
 centrality: the interpretation of it
is that it is in fact the {\em QCD critical temperature}. 

 Fig.\ref{fig_mtspec_all} is an example of a hydro prediction for the
proton $p_t$ spectra. Note that 
no parameters (other than total entropy and EoS) are used,
and the agreement of the predicted shape
is very good, both in normalization and shape.

At
 RHIC 
(rather unexpectedly) we found that  different flow-related slopes
for pion and nucleons 
holds till rather large $p_t\sim 2 \, GeV$, making both
spectra to cross. As a result there are more 
 baryons than pions above this point
(till about
$p_t\sim 5\, GeV$). Observation of that lead to a very good
question:
how far down the spectrum the hydrodynamics should be trusted?

\begin{figure}[t] 
\vspace*{-4mm}
\centerline{\epsfig{file=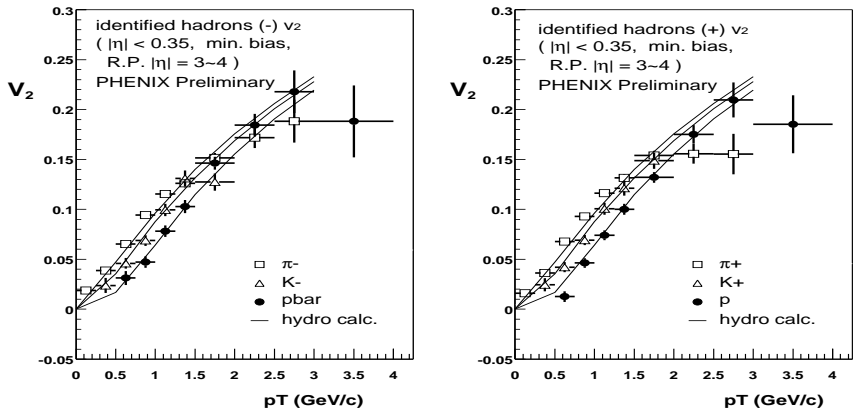,width=12cm,height=6cm}}
\caption{The $p_t$-differential elliptic flow $v_2(p_t)$ from
         minimum bias Au+Au collisions at RHIC, for different 
         identified hadron species (PHENIX). 
with negative (left) and positive 
         (right) charge.
The curves are  hydrodynamic 
         calculations.
\label{fig:v2ptspecies} 
} 
\end{figure} 

\subsection{Elliptic Flow }
\label{sec_Elliptic_Flow}
 
Non-central  heavy ion collisions  produced fireball which has an
almond shape. It would not matter for independently produced secondaries,
but in a collective expansion the shape matters, leading to
 ``elliptic'' flow pattern. 
This is quantified by
$v_i$ harmonics defined as
\be     {dN \over d\phi} = {v_0 \over 2\pi } + {v_2 \over \pi } 
\cos( 2\phi ) +   {v_4 \over \pi }  \cos( 4\phi ) +
                                   \cdots \ee
Each of $v_i$ is a function of centrality (the impact parameter $b$),
rapidity $y$, transverse momentum $p_t$ and, last but not least, the
particle type. By now $v_1,v_2,v_4$ have been studied.
The important feature of elliptic flow is {\em self-quenching},
as a result of which the  elliptic
flow develops earlier than the radial one.
This is why
 it is especially important for understanding the EOS of the QGP.

  The ellipticity depends on a particle mass, again in a predictable
  way
\cite{hydro}. 
Let me show few plots from Kolb and Heinz review \cite{KH_review}
to convince the reader that the elliptic flow is a hydrodynamical
effects.
 
The next Fig.\ref{fig_v2_growth} makes use of one  important fact:
 centrality dependence
of $v_2$ is basically a response
to 
the initial $spatial$ anisotropy of the system, quantified by the parameter
$
   \epsilon=  \langle y^{2}-x^{2}
\rangle /
   \langle y^{2} + x^{2} \rangle ,
$
and so  plotting $v_2/\epsilon$
one basically eliminates the geometry of the
problem and finds all points at some universal curve, see\footnote{The
 horisonal band on this figure marked ``hydro limit'' refers to some
 hydro with ideal gas EoS and simplistic freezeout. It supposed to
 hold at very large entropy density. 
} 
 Fig.\ref{fig_v2_growth}(a). 
 
The main message of this figure is 
that $v_2$  grows  with  multiplicity\footnote{
 This theoretical prediction was made
 at QM99 by Teaney and myself, as well as Kolb and Heinz, prior
to RHIC.}
. The parts (b,c)
of the figure shows how the $v_2$ magnitude was expected to depend
on the collision energy\footnote{Other authors such as Ollitraught
and Heinz et al have used fixed freezeout 
predicted a different energy dependence of $v_2$.}, from Teaney et al \cite{hydro}.

\begin{figure}
\begin{minipage}{8.cm}
\centering
\includegraphics[width=8cm]{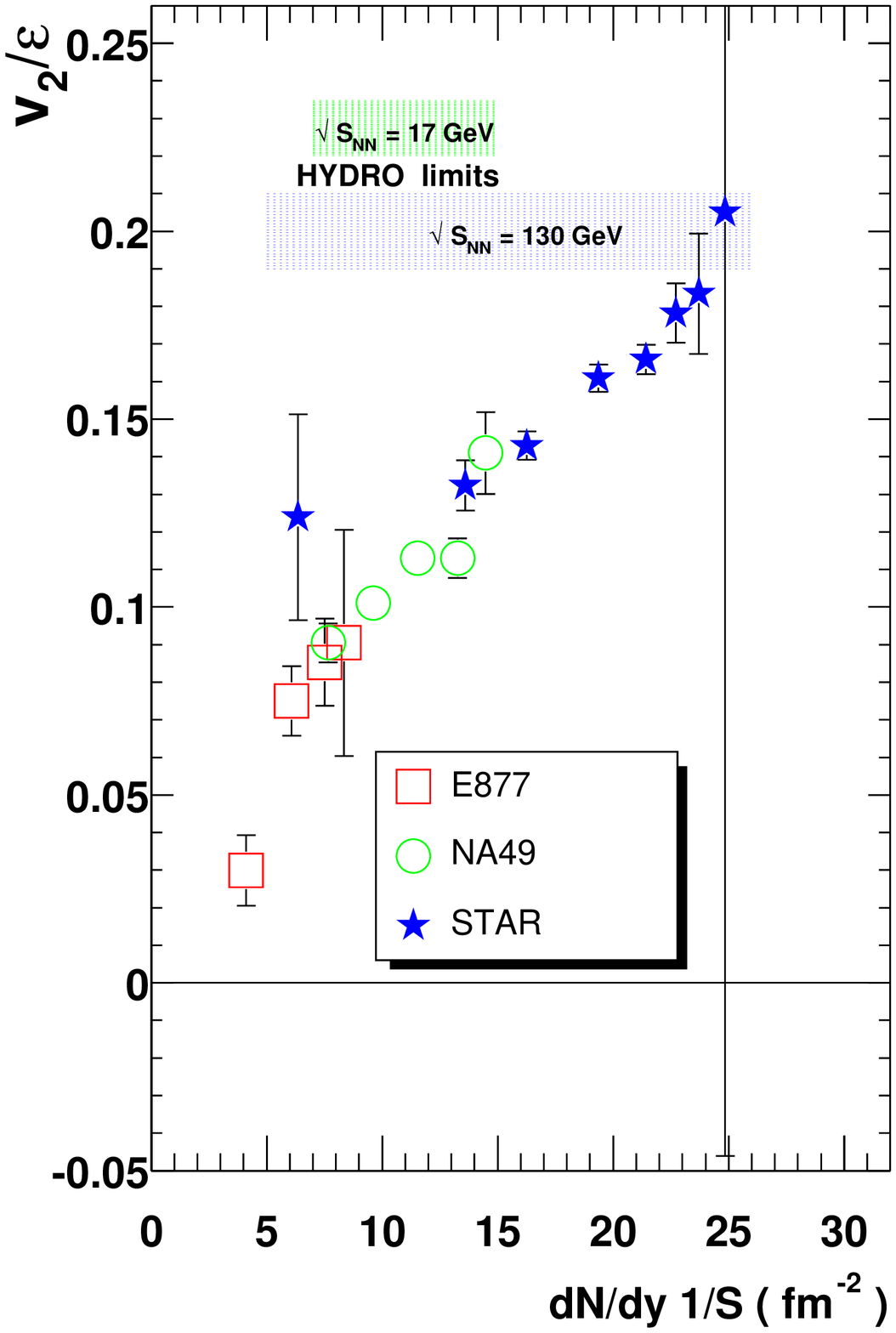}
 \end{minipage} \hspace{1cm}
\begin{minipage}[c]{7.cm}
 \centering 
\includegraphics[width=7.cm]{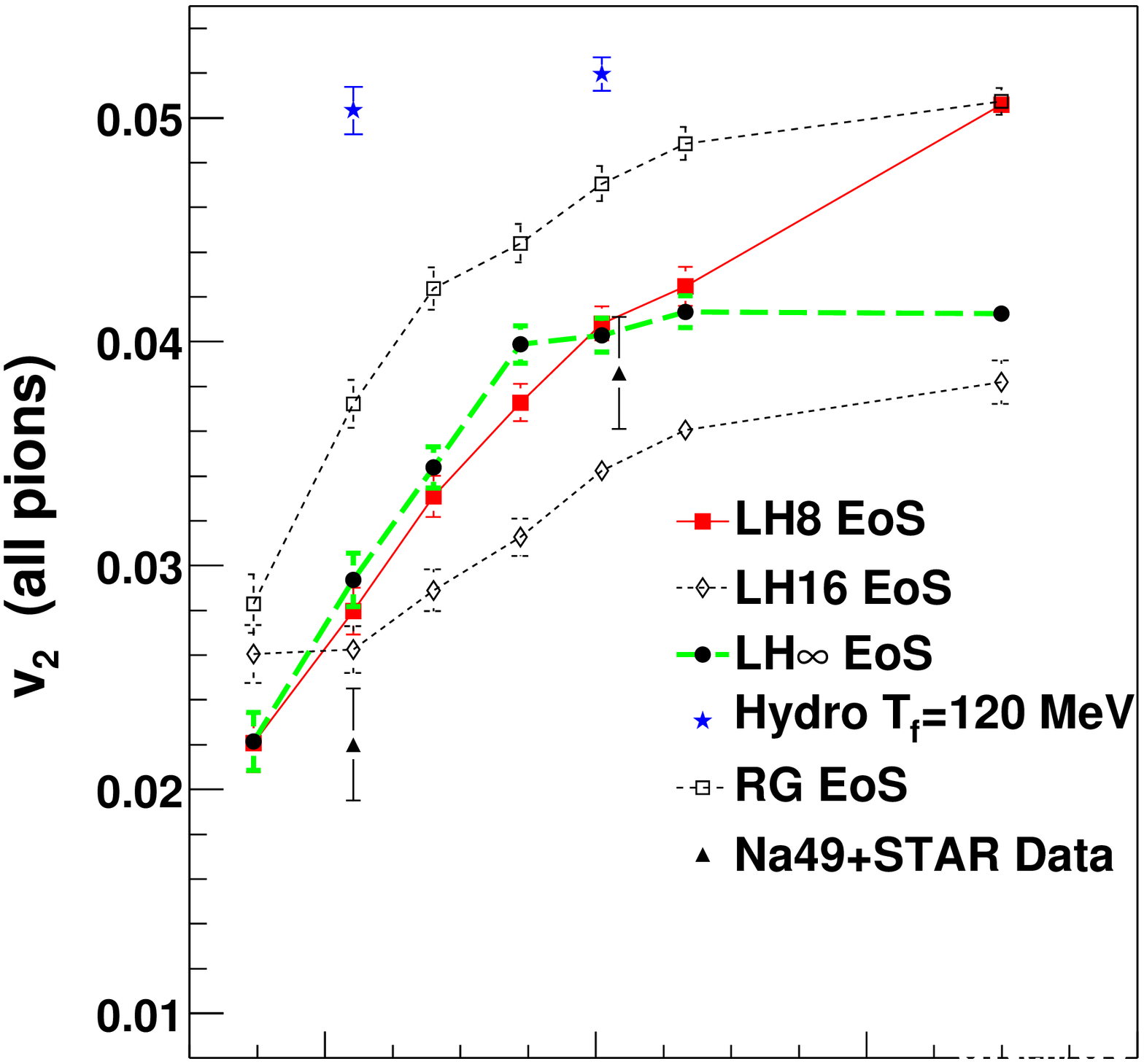}
\includegraphics[width=7.cm]{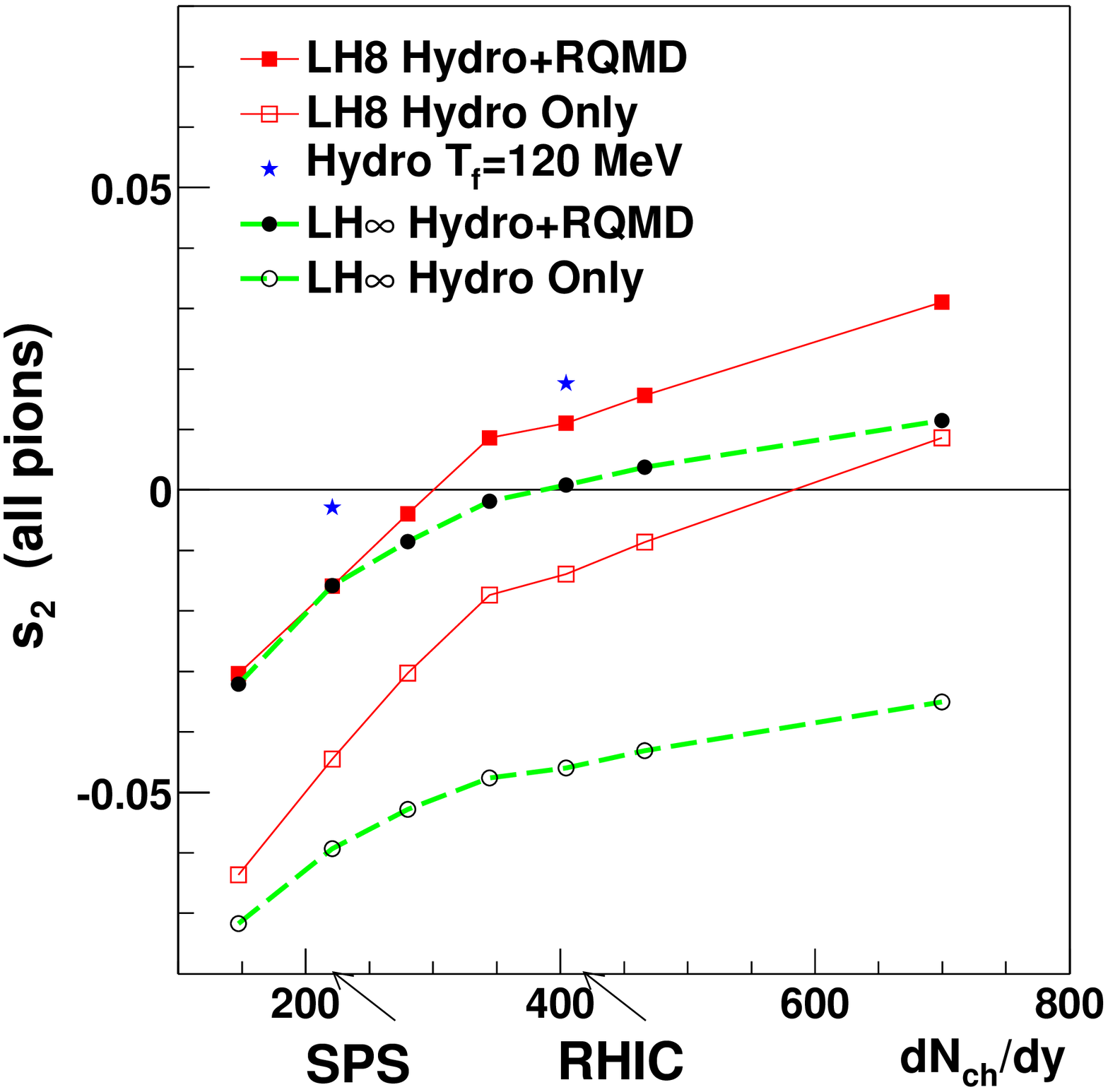} \end{minipage}
\caption{\label{fig_v2_growth}
(a) The compilation of elliptic flow  (the ratio of $v_2/\epsilon$)
dependence on collision energy (represented by the particle multiplicity).  
(b,c) Energy dependence of the elliptic flow predicted by hydro calculation
by Teaney et al \cite{hydro} for different EoS. The
 curve with the latent heat (LH) =800 $MeV/fm^3$ is
the closest to the lattice EoS, and it is also the best fit to
  all
flow data at SPS and RHIC}
 \end{figure}

  We will not have time to discuss details of the
hydrodynamics calculations, which reproduce these data.
 Let me only tell why  RHIC   energy range
is  special. Due to
the QCD  phase transition, the matter is very soft in
the so called ``mixed phase'' energy density region. 
That is why
at SPS energies there was no 
substantial $v_2$ contribution, which only happen at RHIC
due to  ``stiff QGP''.
\subsection{The limits to ideal hydro}
\label{sec_viscous_hydro}
  The ideal  hydrodynamics is $not$ just a bunch of
  conservation laws, but  the local parameterization of the
 stress tensor 
\begin{eqnarray} \label{eqn_tmunu}
T^{\mu \nu} &=& (\epsilon + p) u^{\mu} u^{\nu} - p g^{\mu \nu}   
\end{eqnarray}
Here $\epsilon$ is the energy density, $p$ is the pressure,
 and 
$u^{\mu}=\gamma(1, \bf{v})$ is the proper velocity of the fluid. 

Inclusion of dissipative effects, to the first order
in $l/L$, is possible via the following
  corrections to the  stress  tensor 
 \be \delta T_{\mu\nu}=\eta(\nabla_\mu u_\nu + \nabla_\nu u_\mu
 -{2\over 3}\Delta_{\mu\nu}\nabla_\rho u_\rho)+\xi(\Delta_{\mu\nu}\nabla_\rho u_\rho) 
  \ee
  where the  coefficients  $\eta,\xi$         are called the shear
 and  the  bulk  viscosities.
  In  this  equation  the  following  projection
 operator onto the matter rest frame was used:
  $ \nabla_\mu\equiv\Delta_{\mu\nu}\partial_\nu, \,\,\,
  \Delta_{\mu\nu}\equiv g_{\mu\nu}-u_\mu u_\nu $.
It is further useful to normalize the magnitude of the
viscosity coefficient $\eta$ to the entropy density $s$,
forming a dimensionless ratio. For example
a sound  wave have dispersion law
\be \omega=c_s q-{i\over 2} \q^2 \Gamma_s, \qquad \Gamma_s\equiv
    {4\over 3 T}
{\eta \over s}\ee

\begin{figure}[h]
 \centering 
\includegraphics[width=8.cm]{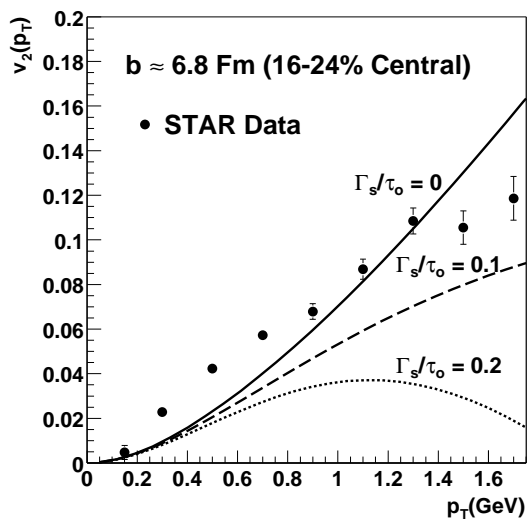}
\caption{ Elliptic flow $v_2$ as a function of $p_T$ for different 
values of $\Gamma_s/\tau_{o}$. The data points 
are four particle cumulants data from the STAR collaboration.
Only statistical errors are shown.
\label{fig_visc_v2_radii}
}
\end{figure}

  Let us now discuss what is the value of QGP  viscosity, following
Teaney \cite{Teaney_visc}. He argued that relative deviations from ideal case
should be $\sim (\eta/s)p_t^2$, and shown that 
 such deviations  are indeed seen in real data.
In  Fig.\ref{fig_visc_v2_radii} we show it for 
the  elliptic flow
parameter $v_2$. Since its value is determined
at sufficiently early times
 -- about 3 fm/c -- the deviation should
 correspond to  the QGP  phase.
The results for different $\Gamma_s/\tau$  shown in 
Fig.\ref{fig_visc_v2_radii}
 deviate from ideal hydro curve 
at $p_\perp\approx 1.6 \, GeV$ which indicates $\Gamma_s/\tau\sim
0.05$ or so. Substituting here the relevant time $\tau\sim 3 fm/c$ we get
 $\Gamma_s\sim .15 fm$. Strong coupling result for typical $T\sim 200
 \, MeV$ at the time gives $\Gamma_s\sim 0.1 fm$, while weak coupling
one would predict much larger value  $\Gamma_s\sim 2 fm$ or so.   
About the same value follows from the gluon cascade with the enhance cross
section by Molnar and Gyulassy \cite{GM}  mentioned above.

\section{Other strongly couples systems}
\subsection{Finite $T$ \N=4 supersymmetric gauge field
  theory at strong coupling} 
   To find a gauge theory in which strong coupling limit
makes sense is a nontrivial task. Furthermore,
to develop tools which would allow a systematic expansion in $inverse$
coupling constant is even more challenging. 
However both problems have been solved during 1990's, first
for a specific -- 4 times supersymmetric -- gauge theory, and then for
   some other examples (we would not discuss).

It is a Conformal Field Theory (CFT), with a non-running coupling.
 Its finite-T version is in the QGP-like
 phase at $any$ coupling, from weak to strong.
As a result of long development by string theorists
based on ``holography'' and ``duality'' ideas, 
the so called AdS/CFT correspondence \cite{Maldacena} was shown,
which states
that {\em CFT in strong coupling is dual to
 a weakly coupled string theory}, albeit in 10 dimensions 
 in a particular gravity field. The finite-T version has a
 gravity metrics with a {\em black hole}, and it is its Howking radiation
which heats up our Universe\footnote{As
Sun worms the Earth.}, represneted by a 4-d surface at some
 distance
from the black hole.  

For a non-string theorist like myself to follow the duality arguments
and gravity-based arguments is a fascinating, near magical experience.
On the other hand, as I am not really interested in string theory and 
consequences of supersymmetry, but
rather in generic effects of any strong coupling,  
it is imperative to find a meaning for gravity-based calculations and results
{\em inside the gauge theory itself}. This work has just started (see
below)
and  we of course have a lot of
problems to solve.

 This CFT  has gluons with $N_c$ colors, which for technical
  reasons
is considered large, 4 types of fermions (gluinos) and 6 scalars. 
The gauge coupling is always combined with the color factor 
 $\lambda\equiv g^2 N_c$, and can be either small or large. 
In the former case one has standard Feynman calculus on a gauge side,
in the latter it is better to used the gravity formulation.
For instance, a  potential between two static quark-like charges
is then described by a string between them, 
which is not straight as in QCD but stretched by gravity into
5-th dimension. The result is a modified Coulomb's 
law for strong coupling ($\lambda\gg 1$)~\cite{Maldacena},
which has the same $r$-dependence but $\sqrt{\lambda}$ instead of
  $\lambda$
\be \label{eqn_new_Coulomb}
V(r)= -{4\pi^2  \over \Gamma(1/4)^4 }{\sqrt{\lambda} \over  r}
\label{coulomb}
\ee
times a very strange coefficient including Euler Gamma function.

 The strong-coupling results for finite $T$  include: \\

(i) bulk  thermodynamics resulted in\cite{thermo} It was found that
the free energy in this limit is
$
{\bf F}(T,N_c,\lambda)=\left((3/4)+ O(1/\lambda^{3/2})\right)
{\bf F} (T,N_c, 0) $
where ${\bf F}(T,N_c,0)\approx N_c^2 T$ is the free (zero coupling) result,
analogous to Stephan-Boltzmann result for blackbody radiation.
 
(ii) the heavy quark potential  is totally screened for
a Debye radius of order $1/T$ \cite{Rey_etal} and leads to
quasiparticle masses of the order $M\sim\sqrt{\lambda}T$\\

(iii)  viscosity of strongly coupled matter was found to be
unusually small, leading to a rather good liquid with hydrodynamical
behavior even at small spatial scales.  In particular,  the viscosity 
to entropy ratio was found to
be~\cite{PSS}
\be 
{\eta \over s }={1\over
 4\pi }
\ee
which is probably the smallest possible value, as it is 
obtained for an infinite coupling.   

  One thing that became clear to us \cite{SZ_cft} 
is the meaning of the black hole. Thinking about strongly coupled
gauge theory one cannot aviod noticing that in particular
partual waves particles  fall at each
other, propagating indefinitely toward small distances. This
for example happens in Klein-Gordon eqn. for $\alpha>1/2$. In
equilibrium, there must also be waved propagating back, from small to
large distances: this constant pair production is a kind of Hawking
radiation.
 
Another is that the ineteraction is transfered by gluons with a
superluminal
speed, $v\sim \lambda^{1/4}>>1$, which may justify potential type
ladder diagrams even for relativistic bound states. 

Ferthermore, as I will try to show, it explained 
 the  puzzle of why thermodynamics can be nearly
independent on the value of the coupling $\lambda$ in strong coupling,
while the composition of matter drastically changes when it changes
 from weak to strong. 
In a naive picture of a quasiparticle gas, one would expect
the Boltzmann factors for quasiparticles to be $exp(-M/T)\sim
exp(-\sqrt{\lambda})\ll 1$, while $p\sim T^4$ clearly demands
the light particles with masses $M\sim T$ at any coupling.  
What those states may be?

 Zahed and myself
\cite{SZ_cft}  proposed an explanation:
these light states are
 deeply bound binary composites, in which the supercritical
Coulomb is  balanced by the centrifugal force.
The argument is rather involved and cannot be given here.
Let me just say that the key is the derivation of the modified
Coulomb law via ladder diagrams  is possible, revealing
that virtual gluons in this regime must fly with 
 super-luminal velocity
$v\approx \lambda^{1/4}\gg 1$. Therefore
 even for relativistically moving 
quasiparticles the interaction can be described by a
potential.
Solving the Klein-Gordon (or Dirac or Yang-Mills) equations for scalars
(or spinors or gluons) in yields {\bf towers of deeply bound states},
extending from large quasiparticle masses $m/T\approx \sqrt(\lambda)$
all the way to small ones $E/T\approx\lambda^0$ that are independent
of the coupling constant. More specifically the spectrum is
\be \label{eqn_spectrum_sctrongcoupling}
E_{nl}=\pm m \left[1+
\left({C\over n+1/2+\sqrt{(l+1/2)^2-C^2} } \right)^2\right]^{-1/2}\,\,.
\ee
\begin{wrapfigure}{l}{6.cm}
\begin{minipage}{5.cm}
\centering
\includegraphics[width=5cm]{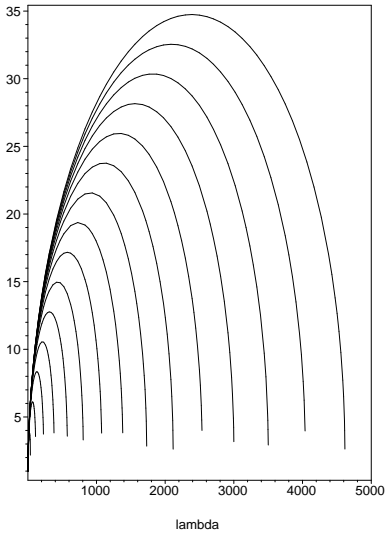}
\caption{\label{fig_wkb} 
The  spectrum of the states versus the 'tHooft coupling constant $\lambda$.
for the levels with fixed $n_r=0$ and the orbital momentum $l=1..15$.
One can see that there are light bound states at any coupling. 
}
\end{minipage}
\end{wrapfigure}

In weak coupling $C=g^2N=\lambda$ is small and the bound states energies
are close to $\pm m$. Specifically~\footnote{The fact that only the
combination $n+l$ appears, i.e. principle  quantum number, is a
consequence of the known Coulomb degeneracy. This is no longer
the case in the relativistic case.}
$
E_{nl}- m\approx -{C^2 m \over 2(n+l+1)^2}\,\,, 
$
which is the known Balmer formulae.
All of that, including the expression above, was known since 
1930's.

New view on this formula,
in the (opposite) strong coupling limit,
 gives the following. If the Coulomb law coefficient is large
$C=(4\pi^2 / \Gamma(1/4)^4 )\sqrt{\lambda} \gg 1$,
the quantized energies are imaginary {\em unless the square
root gets balanced by a sufficiently large angular momentum}. 
 In this regime, one may ignore the $1$ in 
(\ref{eqn_spectrum_sctrongcoupling}) and obtain the {\bf equi-distant}
spectrum of deeply bound states

\be
E_{nl}\approx {m\over C}\left[ 
(n+1/2)+\left((l+1/2)^2-C^2\right)^{1/2} \right] 
\label{wkb}
\ee

 Rather unexpectedly, we have
also found that even though the trajectory of any particular Coulomb bound
state depends critically on the coupling $\lambda$, their average
density remains about $\lambda$-independent constant. This explains
puzzling results obtained using the string theory. Although each level
energy,
and even its existsnce, depend on the coupling, the partition function
is nearly independent of it.

\section{Strongly coupled  trapped atoms}
  If the previous section is too theoretical to some readers,
here is a  table-top experiment. Exciting
  recent development  took place at the frontier
of low temperature physics, with
trapped  $Li^6$ (fermionic) atoms.
 Using
 magnetic field one can use the so called Feshbach
resonances and make a pair of atoms nearly degenerate with their
bound state (usually called a molecule but actually a Cooper pair).
This  results in so large  scattering
length $a$, than a qualitatively new type of matter -- {\em strongly coupled
fermi and bose gases} -- is observed. In particular, this very dilute
systems start
to behave hydrodynamically, displaying elliptic flow very similar to that in
non-central heavy ion collisions.

\begin{wrapfigure}{l}{5.cm}
\begin{minipage}{4.cm}
\centering
\vskip -.4cm
\includegraphics[width=4cm]{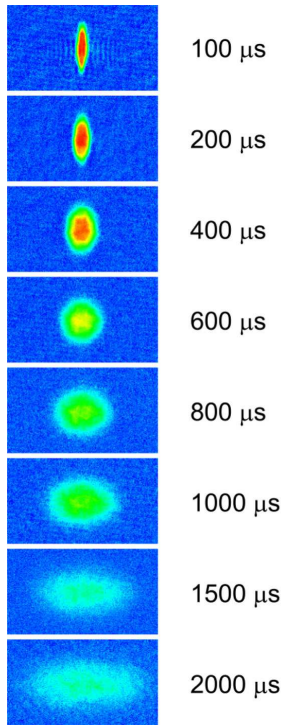}
\caption{\label{fig_atoms} 
Hydrodynamical expansion of trapped $Li^6$, from \cite{Li6}. 
}
\end{minipage}
\end{wrapfigure}

More generally, studies of strongly coupled many-body problems, in
which
the binary scattering length diverges, is being studied in 
at least two other settings: (i) a dilute gas of neutrons, with the
famous virtual level;  (ii)
trapped atomic $Li$ atoms in which the scattering length can be tuned
 till practically infinite values, plus or minus, by applying
a magnetic field which shifts the Feshbach resonances.

Remarkably, for its fermionic version it was indeed
found very recently
that a strong coupling leads to a hydrodynamical behavior
\cite{Li6}. The way it was demonstrated is precisely
the same ``elliptic flow'' as discussed above. One can
start with a deformed trap.  Normally the gas is so dilute
$a n^{1/3} \ll 1$ that atoms just fly away isotropically,
but when tuning to strong coupling regime is done
the expansion is anisotropic and can be described hydrodynamically. 

A number
of other spectacular
 experimental discoveries with trapped $Li^6$
 were also made later. 
 It was found \cite{atoms_capture} that
an adiabatic crossing through the resonance 
 converts nearly all atoms into very loosely bound
(but remarkably stable)  ``Cooper pairs'', which can also
Bose-condense  \cite{atoms_molec_cond} if the temperature is low enough.   
Since in heavy ion collisions the system also crosses
the zero binding lines adiabatically, various bound pairs of quarks and gluons
should also be generated this way. That is probably why we do not
observe
large fluctuations predicted for systems crosseing the QGP-hadronic matter 
boundary.

\section{New picture of QGP, with multiple bound states}

Deconfinement was expected to guarantee
that no hadronic bound states would survive at $T>T_c$, except
perhaps $\bar b b$ states bound by color Coulomb forces.
The earliest suggested QGP signal was
a disappearance of familiar hadronic peaks -- $\rho,\omega,\phi$ mesons --
in the dilepton spectra \cite{Shu_QGP}. Moreover, even small-size deeply-bound
$\bar c c$ states,  $\eta_c,J/\psi$, were expected to melt at
$T\approx T_c$ \cite{MS,KMS}. 
However, as we already mentioned in the Introduction,  there are indications
from the lattice that charmonium and light quarks do create
meson-like states at $T>T_c$.

\begin{figure}
\centering
\includegraphics[width=6.cm]{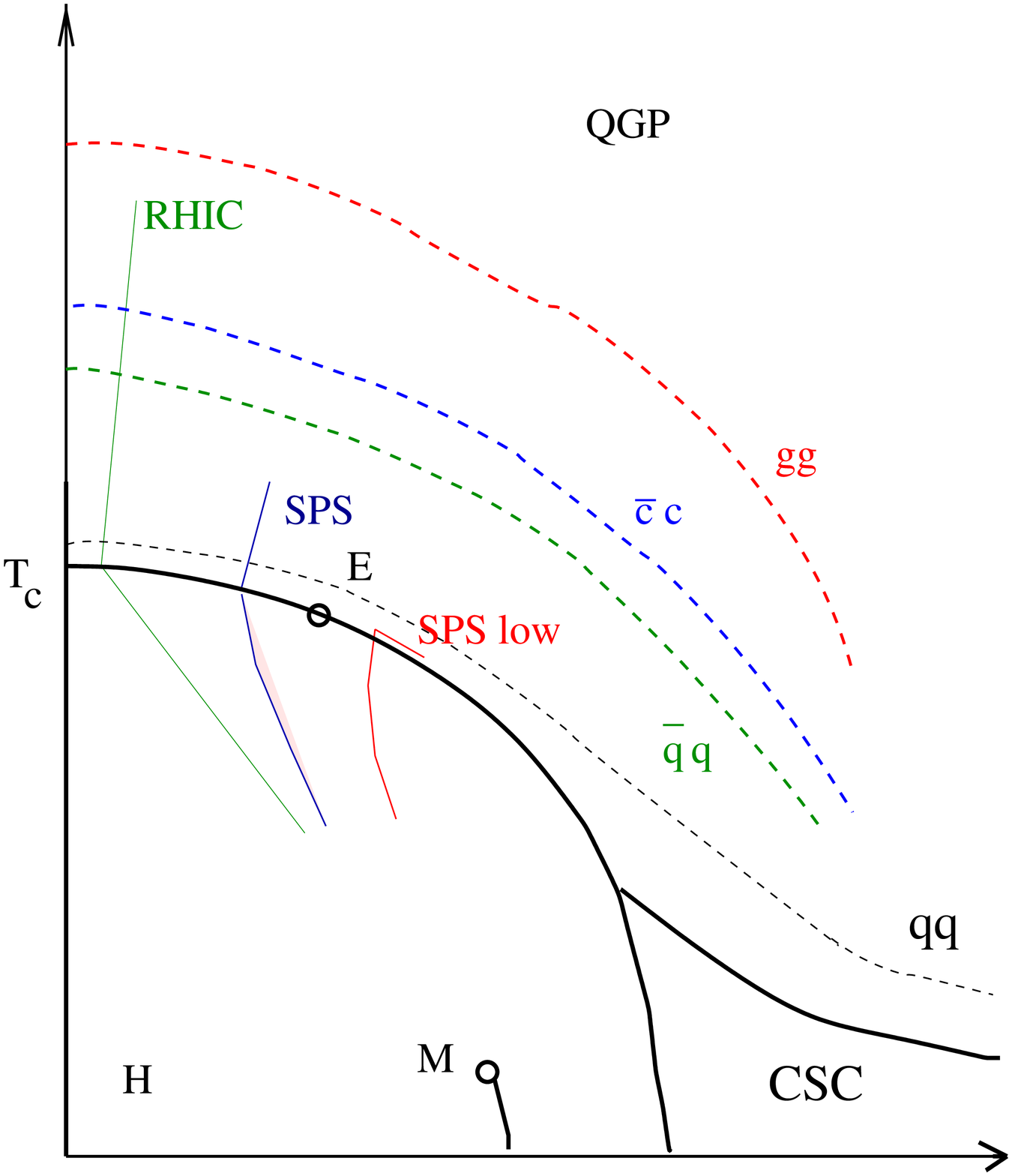}
\includegraphics[width=10.cm]{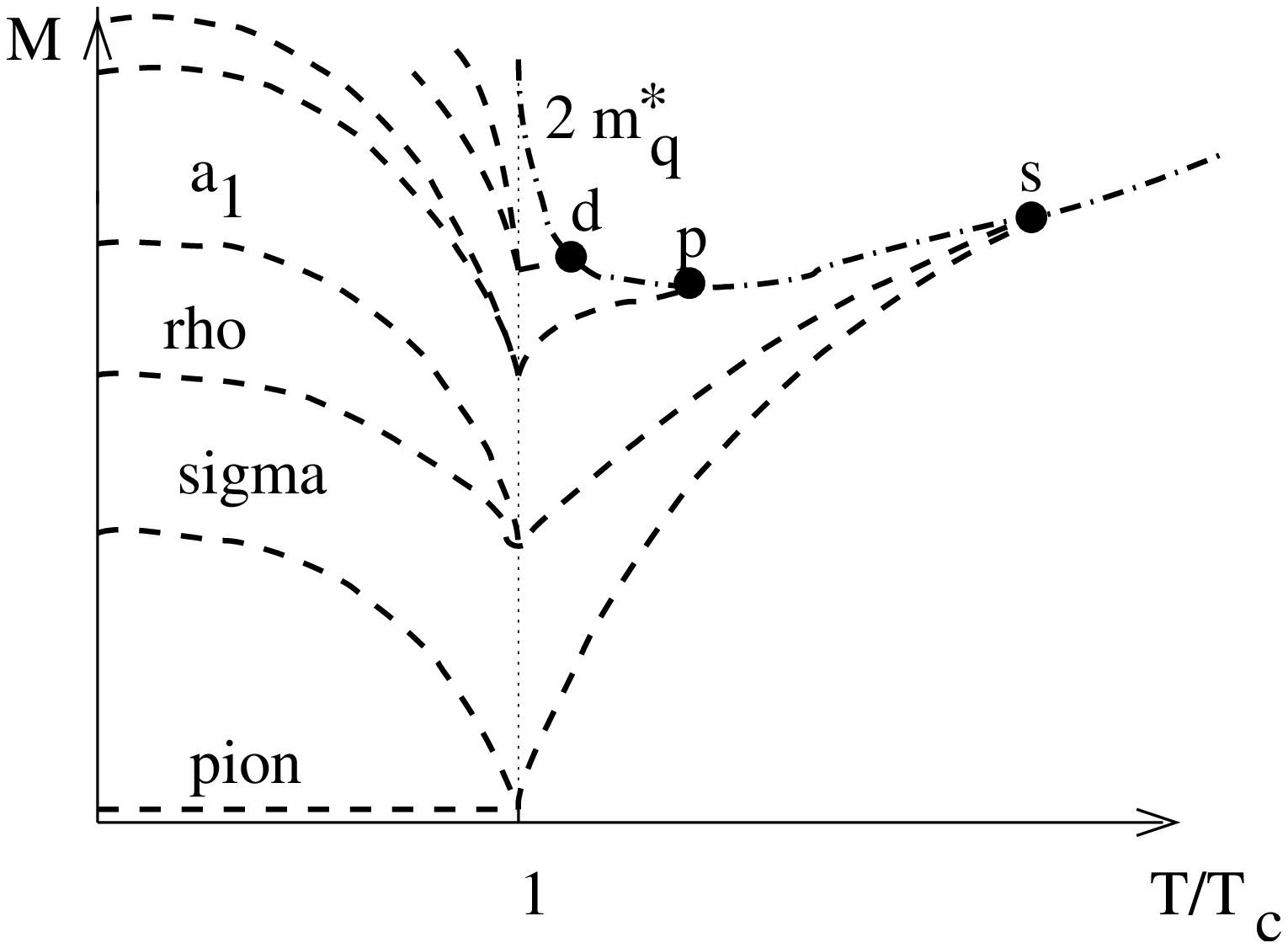}
   \caption{\label{fig_masses_T}
Schematic position of several  zero binding lines on the 
QCD phase diagram (a) and of specific
hadronic masses on temperature $T$ (b). In the latter 
the dash-dotted line shows twice the (chiral) effective mass
of a quark. Black dots marked $s,p,d$ correspond to the points where
the binding vanishes for
states with orbital momentum $l=0,1,2...$.
}
\end{figure}

In the first paper by 
Zahed and myself on the issue\cite{SZ_newqgp} 
we related  presence of
loosely bound pairs of quasiparticles with large rescattering and 
 hydro regime of QGP\footnote{We suggested this mechanism for large
   rescattering before we learned about Feshbach resonances
for atoms, which proves that this mechanism
works.
}.
Indeed, the scattering lengths are supposed to diverge
at 
the {\em zero binding lines} on the phase diagram
 (see Fig.\ref{fig_masses_T}(a)), introduced 
in \cite{SZ_newqgp}. Those line are separate sQGP from wQGP, in which
there are no bound states.

In our  paper \cite{IZ2} we investigate the relationship between four 
(previously disconnected) lattice results: 
{\bf i.} spectral densities from MEM analysis of correlators;
{\bf ii.} static quark free energies $F(R)$; 
{\bf iii.}  quasiparticle masses;
{\bf iv.}  bulk thermodynamics  $p(T)$.
We found  high degree of consistency among  them not known before.
The potentials $V(R)$ derived from $F(R)$ lead
to large number of binary bound states, mostly colored, in
$gq, qq, gg$, on top of the usual $\bar q q$ mesons.
Using the Klein-Gordon equation and ({\bf ii-iii})
we evaluate their binding energies and 
locate the zero binding endpoints on the phase diagram, which
happen to agree with ({\bf i}). We then
estimate the contribution of all states to the bulk thermodynamics
in agreement with ({\bf iv}).  

The bound states of $\bar q q$ can only be colorless mesons
(the octet channel  is repulsive), but in QGP there can
be $colored$ bound states. Quite famous are quark Cooper pairs $qq$
which drive the color superconductivity at sufficiently high density
and low $T$: but pairs themselves should exist outside this region as well.
Gluons can form a number of states with attraction, and there can
also be $gq$ hybrids. A generic reason why we think all of them exist
is that at $T$ close to $T_c$ all quasiparticles are very heavy.
  
  Using a singlet $\bar q q$ as a standard benchmark (the only one studied 
so far on the lattice), one can summarize
the list of all attractive channels in 
the following small Table\ref{tab_all}, indicating the relative strength
of the Coulomb potential and also a number of states.
One can see, there are many hundreds of attractive channels
which can support bound states.

\begin{table}[t]
\centering
\begin{tabular}{llll}\hline
channel & rep. & charge factor & no. of states \\  \hline
$gg$ &  1 & 9/4 & $9_s$ \\
$ gg$ &  8 & 9/8 & $9_s*16$                \\  \hline
$qg+\bar q g$ &  3 & 9/8 & $3_c*6_s*2*N_f$ \\
$qg+\bar q g$ &  6 & 3/8 & $6_c*6_s*2*N_f$        \\  \hline
$\bar q q$ &  1 & 1 & $4_s*N_f^2$  \\  \hline
$qq+\bar q \bar q$  &  3 & 1/2 & $4_s*3_c*2*N_f^2$  \\ \hline
\end{tabular}
\caption{Binary attractive channels discussed in this work, the subscripts
s,c,f mean spin,color and flavor, $N_f=3$ is the number of 
relevant flavors. }
\label{tab_all}
\end{table}


 In another paper, by Brown et al \cite{BLRS}, 
the fate of the  $\bar q q$ bound states
is traced to $T\approx T_c$, where the
Nambu-Goldstone and Wigner-Weyl modes meet.
 The pion binding at total zero mass  is
very difficult to reach from the sQGP side,
and it can only be accomplished by
the combination of (i) the color Coulomb interaction,
(ii) the relativistic effects, 
and (iii) the interaction induced by the instanton-anti-instanton
molecules. The spin-spin forces turned out to be small. While near
$T_{zb}$ all mesons are large-size nonrelativistic objects bound
by Coulomb attraction, near $T_c$ they get much more tightly
bound, with many-body collective interactions becoming important
and making the $\sigma$ and $\pi$ 
masses approach zero (in the chiral limit).
The wave function at the origin grows strongly with binding, and
the near-local four-Fermi interactions induced by the instanton
molecules play an increasingly more important role as the
temperature moves downward toward $T_c$.

\begin{figure}[t]
\centering
\includegraphics[width=7.cm]{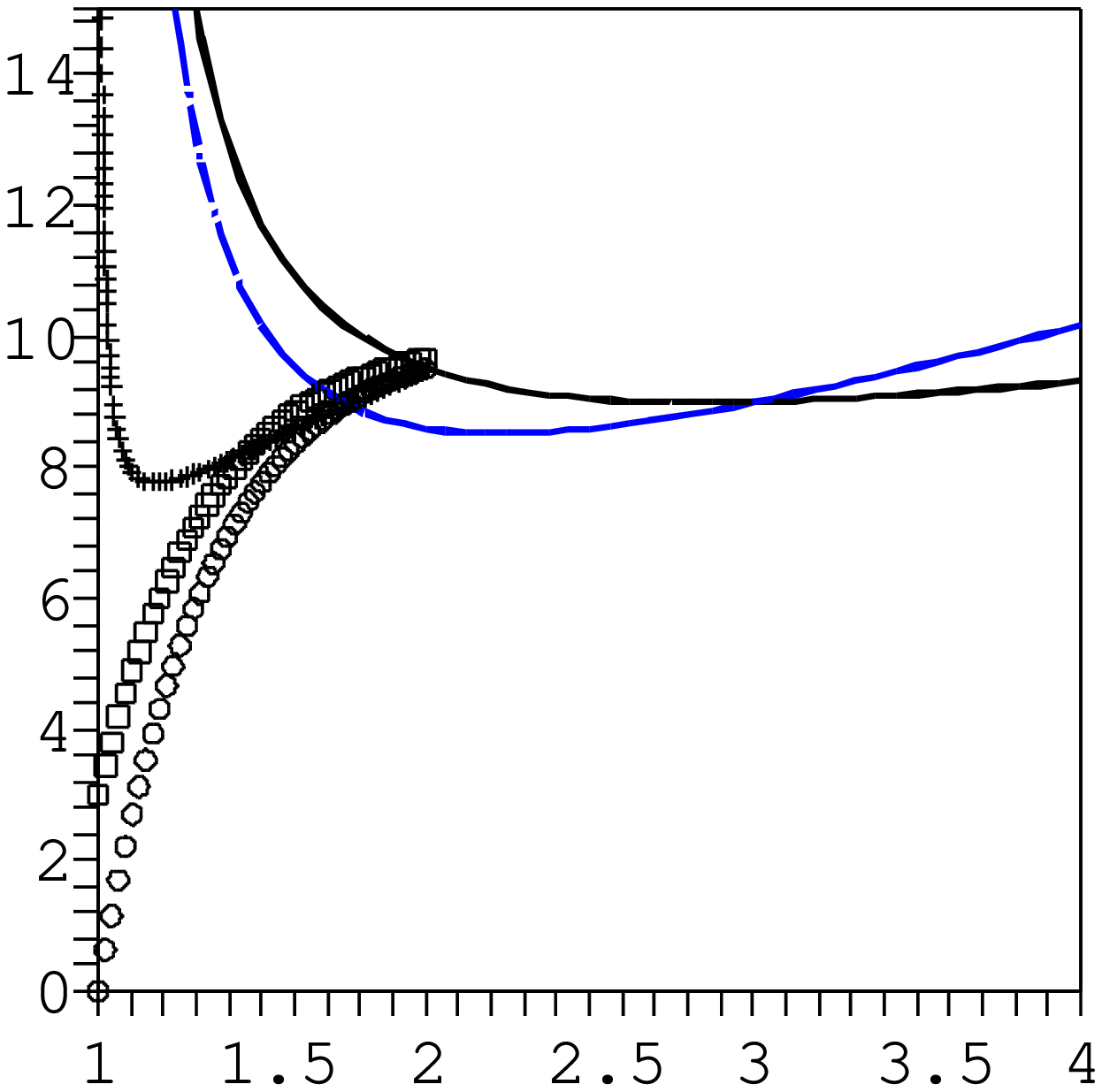}
\includegraphics[width=7.cm]{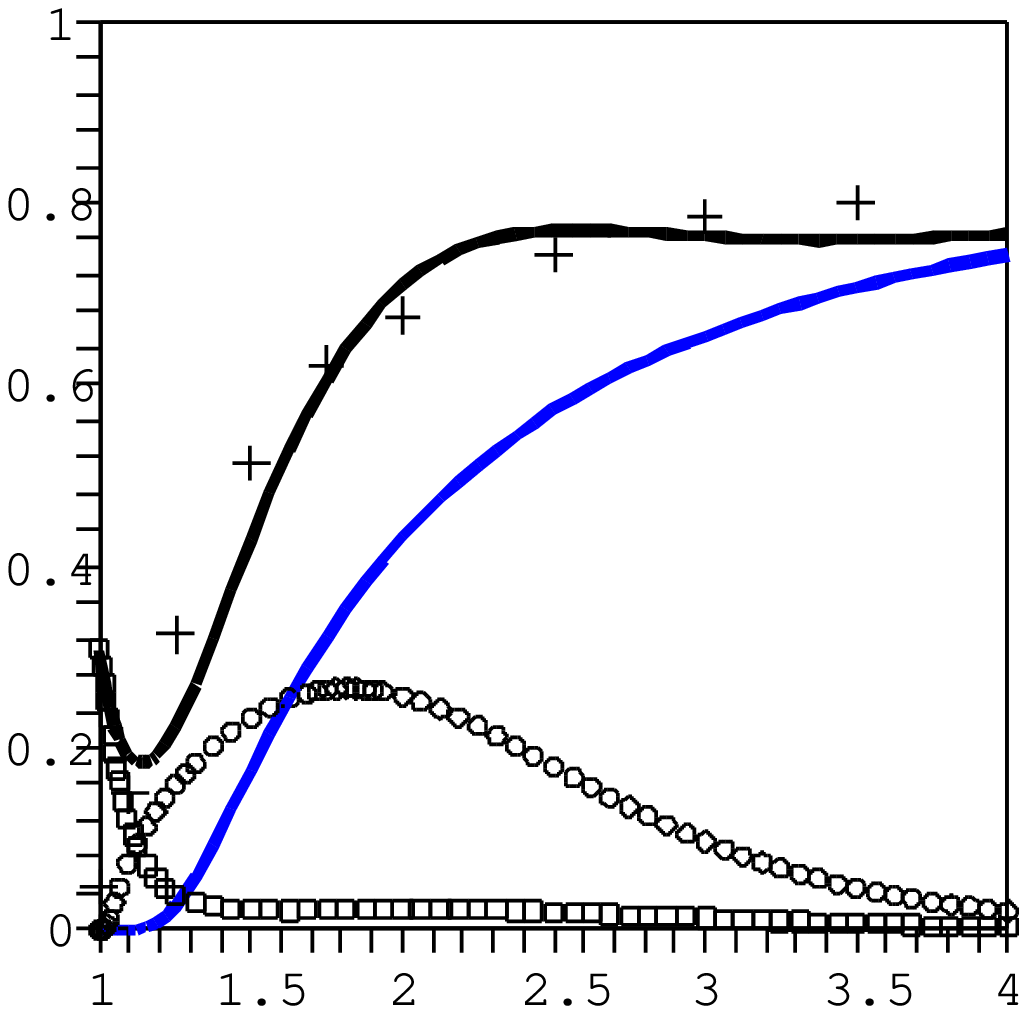}
  \caption{\label{fig_masses}
(a) The lines show twice the effective masses for
quarks and gluons versus temperature $T/T_c$.
Note that for $T<3T_c$ $M_q>M_g$.
Circles and squares indicate estimated masses of the pion-like
and  rho-like $\bar q q$ bound states, while
the crosses stand for 
all colored states.
(b) Pressure (in units of that for a gas of massless and
noninteracting quasiparticles) versus the temperature $T/T_c$.
The crosses  correspond to the $N_f=2$ lattice results, from
Fig.\ref{fig_pressure}(a). The lower solid curve   
is the contribution of unbound quasiparticles, the upper
includes also that of all bound states.
Squares are for  the pion-like
and  rho-like $\bar q q$ bound states combined,
and circles for all the colored bound
states.}
\end{figure}

With all of it included, Zahed and myself \cite{IZ2} had evaluated
the contribution of all these binary bound states into the partition function.
The results for masses of the bound states are shown in
Fig.\ref{fig_masses}(a), and the resulting pressure in
Fig.\ref{fig_masses}(b). We have shown that as the level
closes toward its endpoint, its contribution to pressure 
becomes partially compensated by a repulsive effective interaction
between the unbound quasiparticles. The contribution of the virtual
level above zero quickly disappear. Assembling all these ingredients
together,
we have found that all pieces fit together nicely, reproducing total
pressure
as calculated on the lattice. Hundreds of exotic bound states are
tamed
by small Boltzmann factors, contributing (at RHIC) up to about a half
of the pressure.

Quite recently we have noticed \cite{BSZ} one more important role that
binary bound states may play in sQGP: ``ionization'' of them by
passing jets can contribute to jet quenching. Indeed, in QED
we know it to be dominant at not-too-large gamma factors
$\gamma=1-100$. One motivation is that at SPS the density is not
that much different from RHIC, and yet there is no jet quenching.
One obvious difference with radiative mechanism is that the lost energy is
$not$ to be found in the forward cone.

\section{Brief summary and outlook}
 There is no question that with the RHIC project we got very lucky:
we have not found exactly what we expect, but got instead much more.
 RHIC experiments have shown that
the QGP at $T=(1-2) T_C$ is not a weakly interacting gas of
quasiparticles, being instead a {\em strongly coupled QGP}.

Not only the expected energy density  is
reached, as planned well above the critical region, but we have indeed
created a {\em well-equilibrated matter}, behaving as it should in a bulk.
Surprisingly, this is true not only for most central collisions,
but also for relatively peripheral ones (but not most peripheral,
of course). That means that few hundred
of particles is already enough, even for very specific hydrodynamical
effects like elliptic flow. This is truly surprising: it would not
work for a small drop containing only few hundreds of water molecules! 

What exactly sQGP is we  start to understand only now.
A lot of experience with two other strongly coupled systems was emphasized
in this summary: those are (i) ultra-cold
trapped cold atoms in a large scattering length regime; and 
(ii) the \N=4 supersymmetric Yang-Mills theory, or CFT.
Both show that strong coupling
does indeed lead to a hydrodynamical behavior and small viscosity.
Both teach us a lesson, that small deviation of EoS from ideal gas
does not really mean the matter is weakly interacting. Both show
that resonances between bound and unbound
states seem to play a key role in hydro flow.

 The next notable theoretical achievement is that quantitative
predictions of lattice QCD, on $T_c$ value, the magnitude of the
``latent heat''
and also overall EoS were justified by data. Together
with hydrodynamics (and details like knowledge of the initial
shape of the overlap region and final freezeout conditions)
it provided essentially parameter free theory, which
was exactly on the mark as far as all flows (spectra) are concerned.
(I think the remaining discrepancies with HBT radii will be worked out.)

There are plenty of 
other open questions and even known
experimental puzzles. Is charm floating
with all other flavors, or not? What happens with charmonium, is it enhanced or
depleted?  Why are there so many baryons at
$p_t=2-5 \, GeV$? Why the ellipticity is so large even at very large
$p_t\sim 10 GeV$? Can we see some of the bound states in sQGP in a
dilepton signal? What role the bound state play in jet quenching?

And a super-question is: which part of what we have learned at RHIC
 would still be
applicable
at LHC? Will it still be a bang, or not? Stay tuned...


\begin{thebibliography}{99}
\itemsep -2pt 

\bibitem{Shu_QGP}E. V. Shuryak, 
 {\it Phys.Lett.} {\bf B78} (1978) 150, (full version  {\it Yadernaya Fizika }
 {\bf 28}  (1978) 796),
{\it Phys.Rep.} {\bf 61} (1980) 71. 
 \bibitem{CP_75} J. C. Collins and M J Perry, Phys. Rev Lett. 34(1975)1336
\bibitem{Shu_JETP} E. V. Shuryak, Zh.E.T.F {\bf 74} (1978) 408,
   (Sov. Phys. JETP {\bf 47} (1978) 212),
\bibitem{quenching}
D.~A.~Appel,
Phys.\ Rev.\ D {\bf 33}, 717 (1986).
J.~P.~Blaizot and L.~D.~McLerran,
Phys.\ Rev.\ D {\bf 34}, 2739 (1986).
M.~Gyulassy and M.~Plumer,
Phys.\ Lett.\ B {\bf 243}, 432 (1990).
X.~N.~Wang, M.~Gyulassy and M.~Plumer,
Phys.\ Rev.\ D {\bf 51}, 3436 (1995)
[hep-ph/9408344].


\bibitem{Shu_book2} E.V.Shuryak, The QCD vacuum, hadrons and
  superdense matter. WSPC. 1st edition 1987, 2nd edition 2004. 
\bibitem{Kajantie}K.~Kajantie, M.~Laine, K.~Rummukainen and Y.~Schroder,
Phys.\ Rev.\ D {\bf 67}, 105008 (2003)
[arXiv:hep-ph/0211321].


\bibitem{GM}D.~Molnar and M.~Gyulassy,
Nucl.\ Phys.\ {\bf A697} (2002)  495;
[Erratum-ibid.\ {\bf A703} (2002) 893]

\bibitem{charmonium}
S.~Datta, F.~Karsch, P.~Petreczky and I.~Wetzorke,
{\tt hep-lat/0208012}. 
M. Asakawa and T. Hatsuda,
Nucl. Phys. {\bf A715} (2003) 863c; 
hep-lat/0308034;


\bibitem{Karsch:2002wv}
F.~Karsch et al
Nucl.\ Phys.\  {\bf B715} (2003)  701.
 \bibitem{SZ_radialflow} E. V. Shuryak and 0. V. Zhirov, {\it Phys.Lett.}
 B89 (1979) 253 
 \bibitem{SieRas} P. J. Siemens and J. 0. Rasmussen, 
Phys. Rev Lett. 42(1979)880


\bibitem{Kolb:2002ve}
P.~F.~Kolb and R.~Rapp,
Phys.\ Rev.\ C {\bf 67}, 044903 (2003)
[arXiv:hep-ph/0210222].
\bibitem{hydro}D. Teaney, J. Lauret and E.V.~Shuryak, 
   { Phys. Rev. Lett. }{\bf 86} (2001) 4783, more details in nucl-th/0110037.
  P.F. Kolb, P.Huovinen, U. Heinz, H. Heiselberg,
  { Phys. Lett.} {\bf B500} (2001)  232.
\bibitem{KH_review} P.~F.~Kolb and U.~Heinz,
arXiv:nucl-th/0305084.
\bibitem{Shu_spha} E.V. Shuryak,
 Nucl.Phys.A717:291-321,2003 
\bibitem{Teaney_visc} D.Teaney {\tt nucl-th/0301099}

\bibitem{Maldacena} J.M. Maldacena, 
Adv.Theor.Math.Phys.2:231-252,1998,{\tt hep-th/9711200}.
\bibitem{thermo}
G.T. Horowitz and A. Strominger, { Nucl. Phys.} {\bf B360} (1991) 197.
S.S.Gubser, I.R.Klebanov and A.A. Tseytlin, {Nucl.\ Phys.\ } {\bf B534} (1998) 202

\bibitem{PSS}
G.~Policastro, D.~T.~Son and A.~O.~Starinets,
Phys.\ Rev.\ Lett.\  {\bf 87} (2001) 081601.
\bibitem{Rey_etal} S.-J. Rey, S.Theisen and J.-T. Yee,Nucl.Phys.B527:171-186,1998, {\tt hep-th/9803135}
\bibitem{SZ_cft} E.V. Shuryak and I. Zahed, 
Phys.\ Rev.\ D {\bf 69}, 014011 (2004),
hep-th/0307103.
\bibitem{Li6}K.M.O'Hara et al,
  Science 298,2179, 2002;
 T.Bourdel et al, Phys.Rev.Lett.91 (2003) 020402.

\bibitem{atoms_capture}J.Cubizolles et al, cond-mat/0308018, K.Strecker et
  al,cond-mat/0308318
\bibitem{atoms_molec_cond}M.W. Zwierlein et al,  cond-mat/0311617

\bibitem{MS}
T.~Matsui and H.~Satz,
{Phys.\ Lett.\ } {\bf B178} (1986)  416.
\bibitem{KMS}
F.~Karsch, M.~T.~Mehr and H.~Satz,
Z.\ Phys.\ {\bf C37} (1988)  617.

\bibitem{SZ_newqgp}
E.~V.~Shuryak and I.~Zahed,
arXiv:hep-ph/0307267.
Submitted to PRL
\bibitem{BLRS} G.~E.~Brown, C.~H.~Lee, M.~Rho and E.~Shuryak,
``The anti-q q bound states and instanton molecules at t approx. >= T(C),''
arXiv:hep-ph/0312175.
\bibitem{IZ2}E.~V.~Shuryak and I.~Zahed,Toward the Theory of Binary
  Bound States in Quark-Gluon Plasma, hep-ph/0403127, submitted to PRD.
\bibitem{BSZ}G.~E.~Brown, E.~V.~Shuryak and I.~Zahed,Jet Quenching due
  to Ionization of Binary Bound States in Quark-Gluon Plasma, in progress,

\end{thebibliography}
\end{document}